# Kondo effect with tunable spin orbit interaction in LaTiO₃/CeTiO₃/SrTiO₃ heterostructure

Pramod Ghising, Debarchan Das, Shubhankar Das[†] and Z. Hossain[*]
*Condensed Matter-Low dimensional systems Laboratory, Department of Physics, Indian Institute of Technology, Kanpur-208016, India*

**Abstract**

We have fabricated epitaxial films of CeTiO₃ (CTO) on (001) oriented SrTiO₃ (STO) substrates, which exhibit highly insulating and diamagnetic properties. X-ray photoelectron spectroscopy (XPS) was used to establish the 3+ valence state of the Ce and Ti ions. Furthermore, we have also fabricated δ (CTO) doped LaTiO₃ (LTO)/SrTiO₃ thin films which exhibit variety of interesting properties including Kondo effect and spin orbit interaction (SOI) at low temperatures. The SOI shows a non-monotonic behaviour as the thickness of the CTO layer is increased and is reflected in the value of characteristic SOI field ($B_{SO}$) obtained from weak anti-localization fitting. The maximum value of $B_{SO}$ is 1.00 T for δ layer thickness of 6 u.c. This non-monotonic behaviour of SOI is attributed to the strong screening of the confining potential at the interface. The thicker CTO films in addition to the increased dielectric constant of STO substrate at low temperature leads to stronger screening as a result of which the electrons confined at the interface are spread deeper into the STO bulk where it starts to populate the Ti $d_{xz/yz}$ subbands; consequently the Fermi level crosses over from $d_{xy}$ to the $d_{xz/yz}$ subbands. At the crossover region of $d_{xy} - d_{xz/yz}$ where there is orbital mixing, SOI goes through a maximum.

## 1. Introduction

The interface between perovskite oxides has evoked much interest from the scientific community due to the remarkable phenomena that it exhibits, such as formation of metallic two-dimensional electron gas (2DEG) [1-4], ferromagnetism [5], superconductivity [6][7], enhanced magnetoresistance (MR) [8] and photoconductivity [9-11]. The broken translational symmetry at the interface paves the way for electronic reconstructions which results in the phenomena listed above. One such notable pair of perovskite oxides whose interface has been widely studied is LaTiO₃ (LTO) and SrTiO₃ (STO). LTO is a Mott insulator (Ti³⁺ ions in $d^1$ configuration) with a band gap of 0.2 eV, whereas STO is a band insulator (Ti ions in $d^0$ configuration) with a band gap of 3.2 eV [12][13]. In 2002, Ohtomo *et al* [14] fabricated superlattice of LTO/STO with a sharp interface and reported accumulation of electrons in the Ti-d bands of STO, leading to the formation of 2DEG. The generation of 2DEG at the interface results in a metallic behaviour in spite of the insulating behaviour of bulk LTO and STO. Furthermore, at low temperatures LTO/STO heterostructure exhibits strong spin orbit interaction (SOI) and Kondo scattering [15-17]. Along (001) direction, polar LTO lattice can be thought of as stacked alternating planes of (LaO)⁺ and (TiO₂)⁻, whereas STO structure consists of neutral (SrO)⁰ and (TiO₂)⁰ planes. At the interface of LTO/STO, a potential well is established due to the superposition of Coulomb potentials from the positively charged (LaO)⁺ planes [18]. This potential confines the electrons at the interface resulting in 2DEG. The confined electrons reside on the Ti 3d subbands ($d_{xy}, d_{xz}, d_{yz}$) of the STO [19].

In rare earth titanates (RTiO₃, R = rare earth ions), magnetic structure depends on the f-orbital occupancy. As we go across the lanthanide series from La to Gd, f-orbital electron occupancy increases and R-O-Ti interaction becomes significant in addition to the existing Ti-O-Ti interaction; consequently magnetism in RTiO₃ switches from antiferromagnetic to ferrimagnetic for R= La and Gd, respectively [20]. R=Ce with f¹ electronic configuration presents the case for the onset of R-O-Ti interactions. Bulk CeTiO₃ (CTO) crystallizes in the orthorhombic perovskite structure. The Ce and Ti ions in CTO are trivalent ($d^1$ valence state) and exhibits insulating behaviour [21]. CTO is antiferromagnetic (AF) and neutron diffraction studies have attributed the AF ordering to Ti³⁺ ions, where it was revealed that the AF moments were in a canted state, resulting in a weak ferromagnetic behaviour of the type G$_x$F$_z$ or G$_z$F$_x$ [22].

Similar to polar LTO and LAO, due to the trivalent Ce and Ti ions, CTO exhibits strong prospect for the formation of 2DEG in LTO/STO thin films. The objective of the paper is two-fold: first to grow epitaxial films of CTO on STO substrate and to study its electrical and magnetic properties. Second, since CTO is similar to LTO (La substituted by Ce), δ doping of the LTO/STO interface with CTO (LTO/CTO (δ)/STO) will allow

---

[*]Corresponding author: zakir@iitk.ac.in

[†]Present address: Nano-Magnetism Research Center, Department of Physics, Institute of Nanotechnology and Advanced Materials, Bar-Ilan University, Ramat-Gan 52900, Israel

further insight on the role of the rare-earth ion on the interface phenomena associated with 2DEG. This paper is divided into two section. In Section A we discuss the experimental details and results obtained on CTO/STO film; in section B we discuss the electrical and magnetotransport properties of δ doped LTO/STO films.

**2. Experimental**

Pulsed laser deposition technique was used to fabricate epitaxial thin films of LTO and CTO on $TiO_2$ terminated STO (001) substrate, using KrF excimer laser (Lamda Physik COMPexPro, λ=248 nm). The targets for LTO and CTO were fabricated by standard solid state reaction method. Stoichiometric ratios of $La_2O_3$-$TiO_2$ and $CeO_2$-$TiO_2$ for LTO and CTO, respectively, were ground using a mortar and pestle. The powders were then calcined in a furnace for 12 hours each at 900 ºC, 1000 ºC and 1100ºC. The powders were finally pressed into a pellet and sintered at 1300 ºC for a period of 12 hours. STO substrates were etched in a solution of $NH_4F$, deionized water and HF acid for 30 seconds, to achieve $TiO_2$ termination at its surface. Prior to the deposition, the $TiO_2$ terminated STO (001) substrates was annealed for one hour at 800 ºC in an oxygen pressure of 7.4 x $10^{-2}$ mbar to attain a defect free surface. Furthermore, to achieve layer by layer growth of the film on the substrate, the laser was fired at a repetition rate of 2 Hz and fluence of~ 0.8 $Jcm^{-2}$, which translated to a slow growth rate of 0.10 Å /s. CTO films of thickness 80 Å, 200 Å and 300 Å were deposited on the annealed substrate. The $O_2$ pressure and temperature of the substrate were maintained at $1 \times 10^{-4}$ mbar and 800 ºC, respectively. After deposition, the sample was cooled at the rate of 10 K/min while maintaining the deposition pressure in the chamber. The same deposition conditions were followed for LTO/STO. The viability of these deposition conditions has been well established by our group, and has been used to fabricate excellent epitaxial films of LTO/STO and LAO/STO [10][11][15-17].

For the δ doped LTO/STO samples the deposition procedure is as follows: first, δ unit cells (u.c) of CTO was deposited on the annealed $TiO_2$ terminated STO substrate. Immediately after, 20 u.c of LTO was deposited on top of CTO (δ u.c)/STO. Same deposition procedure (including the temperature and pressure), as used for CTO/STO films were maintained. δ doped LTO films (LTO (20 u.c)/CTO (δ u.c)/STO) with δ=0, 6, 8, 10 u.c were fabricated. X-Ray diffractometer (PANalytical X'Pert PRO), equipped with a Cu-$K_{α1}$ source (λ=1.5405 Å) was used to study the crystalline structure of the deposited films. X-ray photoelectron spectroscopy (XPS) studies was done using PHI 5000 Versa Probe II, FEI Inc. fitted with an Al Kα X-ray source ($hν = 1486.6\ eV$). Magnetization measurements were done in SQUID magnetometer (Cryogenic Limited). The transport properties were measured in a Quantum Design Physical properties measurement system (PPMS). For transport measurements, Ag/Cr electrodes were deposited in the Van der Pauw and four probe geometry using shadow mask. For resistivity and MR measurements electrical contacts on 2 mm x 5 mm samples were made in the four probe geometry. For out-of plane MR, an external magnetic field was applied perpendicular to the sample plane, whereas for in-plane MR, the field was applied along the sample plane and perpendicular to the current direction. Hall measurements were carried out on 3 mm x 3 mm samples in the Van der Pauw geometry (Ag/Cr pads were deposited at the corners for electrical contacts) with an applied external magnetic field, perpendicular to the sample plane.

**3. Results and discussion**

### A. CTO/STO
*A.1 Structure*

Figure 1 shows the results of X-ray diffraction measurement done on 300 Å CTO/STO film. The presence of only (00$2$) film peaks in the $θ − 2θ$ scan (figure 1(a)) confirms the epitaxial growth of CTO on STO (001). From the (002) peak of CTO (figure 1(b)) the film lattice parameter was determined to be 3.846 Å, which is smaller than the pseudocubic lattice parameter ($a_b = 3.939$ Å) calculated from the bulk orthorhombic CTO lattice parameter reported in Ref. 22. Since the lattice parameter of STO ($a_s = 3.905$ Å) is smaller than that of bulk CTO ($a_b$), an in-plane compressive strain acts on CTO films grown on STO. The decrease in out-of plane lattice parameter in spite of the in-plane compressive strain, may be due to defects such as cation non-stoichiometry [23]. XRD-ω scan (figure 1(c)) about the (002) CTO film peak reveals a full width at half maximum (FWHM) of 0.14º. The low FWHM indicates good crystalline structure of the film. The ϕ scan of both the film and the substrate (figure 1(d)) shows four peaks appearing at an interval of 90º which suggests a four-fold symmetry of the CTO film, furthermore the four peaks in the film and the substrate appear at the same value of ϕ confirming a good epitaxial growth of the CTO film on the STO substrate.

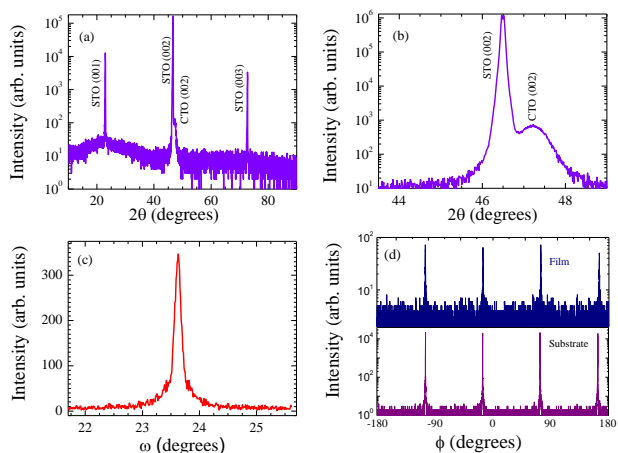

Figure 1. XRD measurement of the CTO film. (a) $θ − 2θ$ scan of the CTO film of 300 Å thickness. (b) The epitaxial growth of the CTO film is confirmed by the (002) XRD peak of the CTO

film. (c) ω scan of the CTO film, the low value of FWHM indicates good crystallinity of the CTO film. (d) ϕ scan about the (022) film and substrate peaks. The set of four coinciding film and substrate peak suggests excellent epitaxial growth.

In order to determine the valence states and stoichiometry of the Ce and Ti in CTO we have carried out XPS measurement on 300 Å CTO/STO film. In order to compensate for the charging effect in the samples, adventitious C 1s peak has been set at 285 eV. The core level XPS spectrum of Ce 3d, Ti 2p and O 1s is presented in figure 2. Ce 3d spectrum (figure 2(a)) shows two spin orbit split peaks, $3d_{5/2}$ and $3d_{3/2}$ at an interval of 18.5 eV. Deconvolution of $3d_{5/2}$ and $3d_{3/2}$ peaks reveals two component each, labelled as ($v_0$, v′) and ($u_0$, u′), respectively. The multipeak splitting of $3d_{5/2}$ and $3d_{3/2}$ occurs due to the final state effect, whereby the core hole potential perturbs the valence electrons and transfer of electrons between Ce 3d and O 2p valence band takes place. The presence of only two components each in $3d_{5/2}$ and $3d_{3/2}$ confirms Ce to be in the 3+ state [24]. However, small quantity of $Ce^{4+}$ is also present as is evident from the small nondescript peak at 916.6 eV which is attributed to $Ce^{4+}$ valence state. Figure 2(b) shows the Ti 2p spin orbit doublet at separation of 5.6 eV. The deconvoluted components of $2p_{3/2}$ at 457.9 eV and 458.6 eV corresponds to $Ti^{3+}$ and $Ti^{4+}$, respectively. The O 1s peak is observed at 529.5 eV (figure 2(c)) with a FWHM of 1.4 eV, the peak at 531.5 is attributed to hydroxyl (OH) species [25].

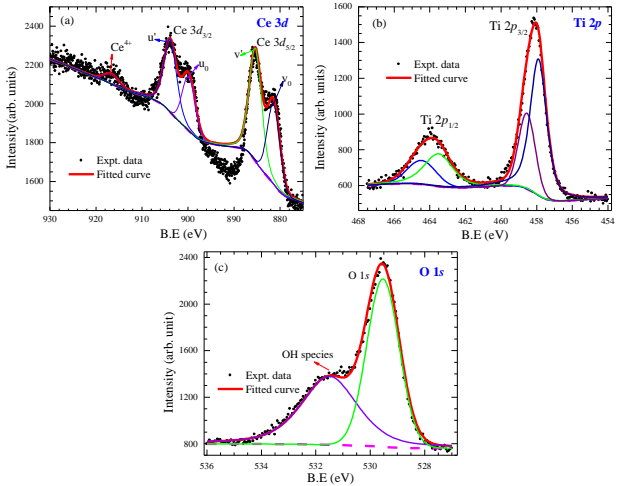

Figure 2. XPS spectrum on 300 Å CTO/STO film, the data was fitted using a combination of Gaussian-Lorentzian peak shapes. (a) Ce 3d spectrum shows the two spin-orbit peaks, $3d_{5/2}$ and $3d_{3/2}$. Deconvolution of the two $3d_{5/2}$ and $3d_{3/2}$ results in two component peaks each, which is ascribed to $Ce^{3+}$ ions. The small peak at 916.6 is due to the $Ce^{4+}$ ions. (b) Ti 2p spectrum with its two spin orbit doublet. The deconvoluted of the Ti 2p peaks reveals the presence of both 4+ and 3+ valence state of Ti. (c) O 1s XPS spectrum.

From the fitting of the XPS data it is seen that both Ce 4+ and Ce 3+ states are present in our sample. However, from the peak intensities, the concentration of $Ce^{4+}$ is negligible as compared to $Ce^{3+}$. From the peak intensities of Ce ($3d_{5/2}$) and Ti ($3d_{3/2}$), we obtain Ce : Ti =0.5:1. This indicates the presence of Ce vacancies in the CTO lattice.

Cation vacancies have been known to result in the oxidation of $Ti^{3+}$ to $Ti^{4+}$ in rare earth titanate ($RTiO_3$) thin films [26][27].We attribute the $Ti^{4+}$ ions in our CTO film to the Ce vacancies ($V_{Ce}$), which is also likely to be responsible for the decrease in lattice parameter observed from the XRD data.

*A.2 Electrical and magnetic properties*

Electrical resistivity measurements were carried out on 80 Å (20 u.c) and 200 Å thick CTO/STO films. The films exhibit very high resistance that could not be measured (the resistance exceeded our measuring instrument's limit∼ 100 $M\Omega$). CTO is a polar perovskite owing to the 3+ valence state of the constituent Ce and Ti ions and conventionally, it is expected to generate 2DEG at the interface when grown on nonpolar STO, leading to metallic behaviour. However, this violation of the general consensus on metallicity of a polar–nonpolar perovskite oxide is not an exception; similar insulating behaviour has been reported for $LaCrO_3$/STO [28] and $LaMnO_3$/STO [29]. Our experimental results corroborate that polar/nonpolar interface is not the only factor responsible for the origin of the interfacial 2DEG. For instance, in $NdTiO_3$/STO films, it has been reported that a single $Nd^{3+}$ ion vacancy generates three holes, which localizes around three $Ti^{3+}$ ions, resulting in a single polaron in $Ti^{4+}$ state [26]. This oxidation of $Ti^{3+}$ to $Ti^{4+}$ severely depletes the electrons available for transfer from polar NTO to the STO interface. We cite similar mechanism for the lack of 2DEG in our CTO/STO thin film, owing to $V_{Ce}$.

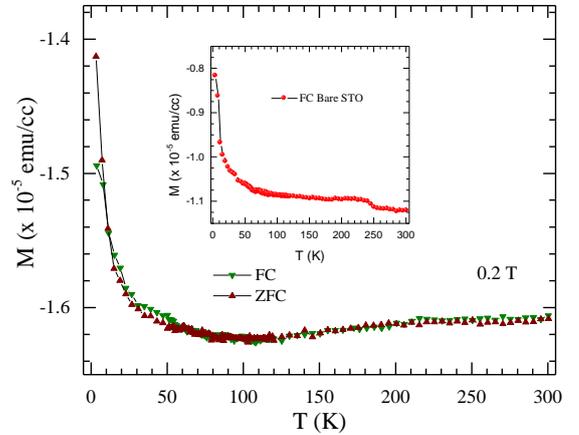

Figure 3. FC and ZFC magnetization data of 300 Å CTO/STO. The data were taken during heating part of the measurement under an applied field of 0.2 T. The sample was field cooled in field of 0.2 T during the FC measurement. The inset shows the FC magnetization of bare STO substrate under the same applied filed of 0.2 T. The Kink in the MT curve of STO at 250 K may be due to the impurities in the STO.

Magnetization measurement as a function of temperature M(T), was performed on a 300 Å CTO/STO film. The measurements were carried out at an applied field of 0.2 Tesla under zero field cooled (ZFC) and field cooled (FC) conditions. The results are plotted in figure 3. From M(T) data, a diamagnetic behaviour is observed in the CTO film as opposed to the antiferromagnetic behaviour observed in bulk CTO. Antiferromagnetism in bulk CTO is attributed to the unpaired $Ti^{3+}$ $d^1$ electrons. In contrast, the emergence of diamagnetism in CTO film suggests the oxidation of $Ti^{3+}$ to $Ti^{4+}$ with completely empty $d^0$ bands i.e. absence of unpaired electrons, all the lower orbitals are completely filled. The STO substrate with empty $Ti^{4+}$ $d^0$ orbital is known to exhibit diamagnetism [30][31]. For reference we also carried out M(T) measurement on a bare $TiO_2$ terminated STO substrate, subjected to the same pressure and temperature conditions used during deposition of CTO film. The temperature dependent FC magnetization profile of the bare STO substrate is shown in the inset of figure 3. The M(T) curve of the STO substrate shows a similar diamagnetic behaviour as that of CTO film, however, strength of the diamagnetic signal of the CTO film is enhanced by nearly a factor of 2.

**B. δ (CTO) - doped LTO/STO**

*B.1. Electrical Transport*

Figure 4 (a) shows the temperature dependent sheet resistance $R_\square(T)$ of the LTO (20 u.c)/CTO (δ u.c)/STO samples with δ=0, 6, 10 u.c. Metallic behaviour in all the three samples is evident from the decreasing $R_\square(T)$ as a function of temperature. However, the resistance of the sample increases with δ layer thickness. Das *et al* [15] have reported a similar $R_\square(T)$ behaviour on $LaCrO_3$ (δ layer) doped LTO/STO system. Using electron energy loss spectroscopy (EELS), they observed that the $LaCrO_3$ δ layer traps the electrons that are being transferred from the LTO to the STO side, which enhances $R_\square(T)$ [15]. We anticipate a similar mechanism for the enhancement in $R_\square(T)$ for our δ doped samples. Hall measurement performed on these samples at 300 K and 2 K reveals an inverse relation of the interface carrier concentration ($n_e$) with the δ layer thickness (insets figure 5 (a) and (b)). Thus, the electrons transferred from the LTO to STO due to polar reconstruction is absorbed by the CTO layer, which results in enhanced $R_\square(T)$. This picture also lends strong credibility to polar reconstruction being responsible for the formation of 2DEG at the LTO/STO interface.

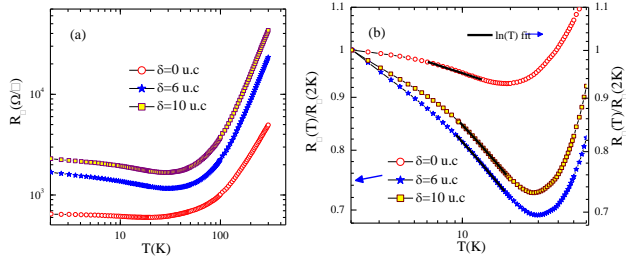

Figure 4. (a) Sheet resistance of the LTO/STO and CTO doped LTO/STO film (with δ layer thickness of 6 and 10 u.c) as a function of temperature. At lower temperatures the resistance of the films display an upturn followed by saturation. (b) To properly emphasize the low temperature upturn in the resistance, $R_\square$ has been normalized by the value of $R_\square$ at T=2K. The black solid line represents $ln(T)$ fit to the $R_\square(T)$ data.

Another notable feature of the $R_\square(T)$ data is the resistance minimum in the temperature range of 20-30 K. In order to emphasize the resistance minima followed by the upturn, we have plotted $R_\square(T)/R_\square(T=2 K)$ vs T in figure 4(b). Below the resistance minimum temperature ($T_{min}$), $R_\square$ gradually increases and tends to saturate below 10 K. A fit to the experimental data below $T_{min}$ exhibits $ln(T)$ behaviour (black solid line in figure 4(b)). The $ln(T)$ rise in $R_\square(T)$ at low temperatures could be associated with a number of quantum scattering phenomena viz. weak localization (WL) [32]-[34], Kondo scattering [15]-[17] and electron-electron interaction (EEI) [35-39]. The origin of WL is purely quantum mechanical, which occurs due to the diffusive motion of electron in disordered system. The time reversed partial waves of the electrons undergo constructive interference, which increases the probability of backscattering and hence a resultant increase in resistance of the material results. On the other hand, Kondo scattering is prominent in the low temperature region where the scattering due to phonons is minimal. Magnetic impurities in the lattice interact with the electrons in the system which leads to an anti-parallel alignment of the spins between the electrons and the magnetic impurity ions. At low temperatures the contribution of Kondo scattering to the electrical resistivity is proportional to $ln(T)$. As the temperature is further lowered the spin of the impurity ions is screened by the spins of the conductions electrons, and below a characteristic temperature (Kondo Temperature ($T_K$)), the resistance gradually saturates. Similarly, EEI also introduces a logarithmic correction to the resistance [38][40][41].

In figure 4(b), at low temperatures (T << $T_{min}$) $R_\square(T)$ goes to a saturation. This behaviour of $R_\square(T)$ suggests the presence of one more scattering phenomena, the weak antilocalization (WAL) [42-45]. WAL effect occurs at low temperatures, in the presence of SOI which introduces a phase difference between the backscattered electron partial waves, leading to a destructive interference; this results in reduced backscattering of electrons and hence a saturating behaviour in $R_\square(T)$.

*B.2 Magneto Transport*

In order to understand the mechanism behind the resistance upturn below $T_{min}$, we study the magneto transport properties of the δ=0, 6, 10 u.c samples at 2, 5 and 20 K. To determine the charge carrier density we have performed Hall measurement on these samples at 2 and 300 K. The results of Hall measurement are plotted in figure 5. The negative slope of $R_H$ $Vs$ $B$ plot suggests the majority charge carriers at the interface are electrons. The insets of figure 5 (a) and (b) show the charge carrier density as a function of δ layer thickness at 300 K and 2 K, respectively. It is observed that $n_e$ has an inverse relation with δ layer thickness at both temperatures. At 300 K, $R_H - B$ plots for the $\delta = 6$ and $10$ u.c exhibit non-linear behaviour at low fields. Kim *et al* [46] have observed similar non-linear behaviour in the $R_H - B$ data of LTO/STO superlattice, which occurs due to multiband occupancy of the charge carriers.

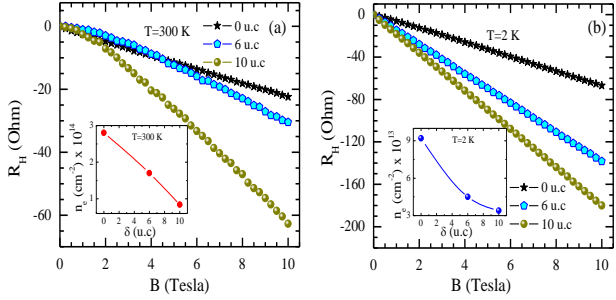

Figure 5. Hall resistance measured as a function of magnetic field for various $\delta$ doping thickness at (a) 300 K and (b) 2 K. The negative slope indicates the charge carriers to be electrons. Carrier density as a function of δ layer thickness at 300 K and 2 K is shown in the inset of 5 (a) and (b), respectively. Inverse relation between charge carrier density and δ layer thickness is observed.

In figure 6 we show the plot of MR data with the magnetic field applied perpendicular to the sample plane. We define, $MR\% = \frac{R(B)-R(0)}{R(0)} \times 100$, where R(B) is the resistance at an applied field $B$ and R(0) is the resistance at zero field. All the three samples display positive MR.

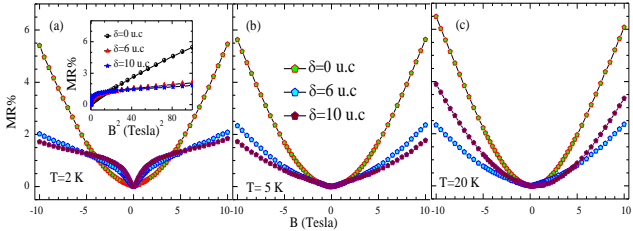

Figure 6. (a)-(c) Out of plane MR of the LTO/CTO(δ u.c)/STO films for δ=0,6,10 at 2, 5 and 10K, respectively. The sharp cusp in the MR data for δ=6,10 u.c samples at 2K is ascribed to the WAL effect. The inset in (a) is the Kohler plot at 2K. The linear behaviour of the Kohler's plot at higher fields indicates the presence classical orbital MR.

External magnetic field has profound impact on WL, WAL and EEI, and can be used as a tool to distinguish between them. Backscattering is suppressed in the presence of a magnetic field which introduces a phase difference between the time reversed electron partial waves and destroys the constructive interference. Thus WL exhibits negative MR. Whereas for WAL the magnetic field destroys the existing destructive interference [47] and hence leads to positive MR. In the case of EEI, Zeeman splitting by an amount of $g\mu_B B$ [48][49] leads to a decrease in conductivity. Therefore magnetic field leads to a negative magnetoconductivity in case of EEI and in the 2D limit, it is given by [47]-

$$\Delta\sigma(B) = -\frac{e^2 F}{4\pi^2 \hbar} A\left(\frac{g\mu_B B}{k_B T}\right) \qquad 1(a)$$

$$A\left(\frac{g\mu_B B}{k_B T}\right) = \begin{cases} \ln\left(\frac{1}{1.3}\frac{g\mu_B B}{k_B T}\right), & \frac{g\mu_B B}{k_B T} \gg 1 \\ 0.084 \left(\frac{g\mu_B B}{k_B T}\right)^2, & \frac{g\mu_B B}{k_B T} \ll 1 \end{cases} \qquad 1(b)$$

Where, $\Delta\sigma(B) = \sigma(B) - \sigma(0)$, $F$ is the electron screening factor, $k_B$ is the Boltzmann constant, $g$ is the electron g-factor and $\mu_B$ is the Bohr magneton. The positive MR at perpendicular fields rule out WL as the mechanism for scattering below $T_{min}$. Therefore, the observed positive parabolic MR in our samples seems to suggest a contribution from EEI or classical orbital effect. At 2 K, the maximum limit of B for parabolic contribution to MR from EEI is $\sim 1.5\,T$ (eqn. 1(b)). Thus, $B^2$ dependence of MR at higher fields suggests the contribution of classical MR as well. MR due to classical orbital MR follows Kohler's rule, $\Delta R/R_\square(0) = k(B/R_\square(0))^2$ [51-53]. Kohler's plot for the doped samples at 2K is shown in the inset of figure 6(a). All the samples exhibit linear behaviour at higher fields, which indicates the presence of classical MR. Figure 7 (a) shows the classical $B^2$ fit to the R-H data for δ=0 u.c sample at 2 K. It is seen that the $B^2$ dependence provides a good fit to the experimental data for $B > 6$ T only, and at $B = 0$ T the $B^2$ fit gives a value of $R_\square(B = 0)$ greater than that obtained experimentally (Fig. 7 (a)). The deviation of data from $B^2$ fit at low field and low temperature may be attributed to WAL which reduces $R_\square(B = 0)$; for $B \neq 0$ WAL is suppressed and $R_\square(B)$ increases. A theoretical model developed by Hikami, Larkin and Nagaoka (HLN) gives an account of the WAL contribution to MR and for 2D systems and is expressed as[17][32][42][54-56]-

$$\begin{aligned}\Delta\sigma(B) &= -\frac{e^2}{2\pi^2\hbar}\Big[\frac{1}{2}\Psi\left(\frac{1}{2}+\frac{B_\varphi}{B}\right) \\ &\quad -\frac{1}{2}\ln\left(\frac{B_\varphi}{B}\right) - \Psi\left(\frac{1}{2}+\frac{B_\varphi+B_{SO}}{B}\right) + \ln\left(\frac{B_\varphi+B_{SO}}{B}\right) \\ &\quad -\frac{1}{2}\Psi\left(\frac{1}{2}+\frac{B_\varphi+2B_{SO}}{B}\right) + \frac{1}{2}\ln\left(\frac{B_\varphi+2B_{SO}}{B}\right)\Big]\end{aligned} \qquad (2)$$

Where, $\Psi(x)$ is the digamma function, $B_\varphi = \hbar/{4eL_\varphi^2}$ is the dephasing field with $L_\varphi$ as the effective dephasing length, $B_{SO} = \hbar/{4eL_{SO}^2}$ is the characteristic field related to SOI; $L_{SO}$ is the spin relaxation length. HLN theory only takes $k$-cubic SOI into account.

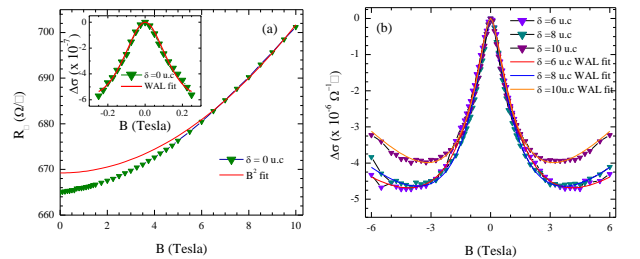

Figure. 7. (a) $R_\square(B)$ data for δ=0 film at 2 K. The experimental data follows a $B^2$ behaviour for B > 5 T, the red solid line represents the $B^2$ fit to the experimental data, the deviation from the $B^2$ fit at lower fields is attributed to WAL. The inset shows WAL fit to MR data in the range -0.3 < B < 0.3 T using Eqn. (2). (b) WAL fit of MR data using Eqn. (2), for δ=6, 8 and 10 u.c samples at 2 K, respectively. The $B^2$ background has been subtracted.

The experimental data in the lower field range ($-0.3 < B < 0.3$ T), can be fitted to a good precision using WAL effect (eqn. 2). This fitting is shown in the inset of figure 7 (a) for δ=0. The MR plot at 2 K (figure 6(a)) for δ=6 and 10 u.c shows a cusp at lower field values. Such cusp in MR data has been observed in previous studies [15][17][38][42][45], and is attributed to the WAL effect. In order to verify the WAL contribution, eqn. (2) is used to fit the experimental data (after subtracting the classical Lorentz $B^2$ background) for the low field region ($-6 < B < 6$ T). Excellent fitting of experimental data (figure 7 (b)) with eqn. (2) justifies the presence of WAL, and it is consistent with our $R_\square(T)$ data which exhibits a saturating tendency in this temperature range. The fitting parameters are shown in Table I. $B_{SO}$, which is a measure of the strength of SOI shows a non-monotonic behaviour with δ (or $n_e$), it is maximum for $\delta = 6$ u.c and decreases for $\delta = 0, 8$ and $10$ u.c. This observation generates new possibilities for tuning the SOI with δ doping.

One more scattering mechanism which can be responsible for the low temperature resistance minima, followed by a $\ln(T)$ increase and finally a saturation tendency in the $R_\square(T)$ curve is Kondo scattering. We further perform in-plane MR measurement to probe it's existence. The in-plane MR ($MR_\parallel$) data for δ=0, 6, 10 u.c samples at T=2 and 10 K is shown in figure 8 (a), (b). It is observed that $MR_\parallel$ increases with δ layer thickness. At 10 K all the three samples exhibit negative MR. At 2 K,

Table I: Fitting parameters obtained from WAL fit of out-of plane MR data at 2 K, for δ=0, 6, 8 and 10 u.c samples.

| Sample | $B_\varphi$(T) | $B_{SO}$(T) |
|---|---|---|
| δ=0 u.c | 0.07 | 0.09 |
| δ=6 u.c | 0.19 | 1.00 |
| δ=8 u.c | 0.16 | 0.89 |
| δ=10 u.c | 0.16 | 0.78 |

δ=0 u.c sample exhibits negative MR, whereas δ=6 and 10 u.c samples exhibit an interesting MR behaviour: the MR of these δ doped samples show a cusp around B=0 T, with a positive maximum at 4 T and as the field is increased beyond 7 T the MR becomes negative. This behaviour in MR suggests the presence of more than one scattering mechanisms in δ doped samples at 2 K. Similar cusp in in-plane MR has also been observed in LAO/STO [57], Bi [58], Mg [59] and topological insulators [39][45] and was attributed to WAL effect. On the other hand, magnetothermopower measurements have unambiguously confirmed the presence of Kondo scattering in LTO/STO thin films [16]. In our thin film samples, the $\ln(T)$ behaviour of the $R_\square(T)$ below $T_{min}$ coupled with the observed negative $MR_\parallel$ is strongly suggestive of Kondo scattering. Thus the experimental observations point to the presence of both Kondo and WAL effect in the δ doped samples at 2 K. To confirm Kondo scattering, we fit our experimental observations with a simple Kondo model [60]:

$$R^{model} = R_0 + R_K(B_\parallel/B_1) \qquad (3\,a)$$

Where, $R_0$ is the residual resistance, $B_1$ is a field scale related to the Kondo temperature ($T_K$) and g-factor of the magnetic impurity spin and $R_K(B_\parallel/B_1)$ is the magnetoresistance of the Kondo impurity at zero temperature and is given by:

$$R_K(B_\parallel/B_1) = R_K(B_\parallel = 0)cos^2\left(\frac{\pi}{2}M(B_\parallel/B_1)\right) \quad (3\,b)$$

$M(B_\parallel/B_1)$ is the magnetization of the Kondo impurity at zero temperature and is expressed as:

$$M(B_\parallel/B_1) =$$
$$\begin{cases} \frac{1}{\sqrt{2\pi}}\sum_{k=0}^\infty \left(-\frac{1}{2}\right)^k (k!)^{-1}\left(k+\frac{1}{2}\right)^{\left(k-\frac{1}{2}\right)} e^{-\left(k+\frac{1}{2}\right)}\left(\frac{B_\parallel}{B_1}\right)^{2k+1}, \\ \qquad\qquad\qquad B_\parallel \leq \sqrt{2}B_1 \\ 1 - \pi^{-3/2}\int \frac{dt}{t}\sin(\pi t)\,e^{-t\ln\left(\frac{t}{2e}\right)}\left(\frac{B_\parallel}{B_1}\right)\Gamma\left(t+\frac{1}{2}\right), B_\parallel \geq \sqrt{2}B_1 \end{cases}$$

$$(3c)$$

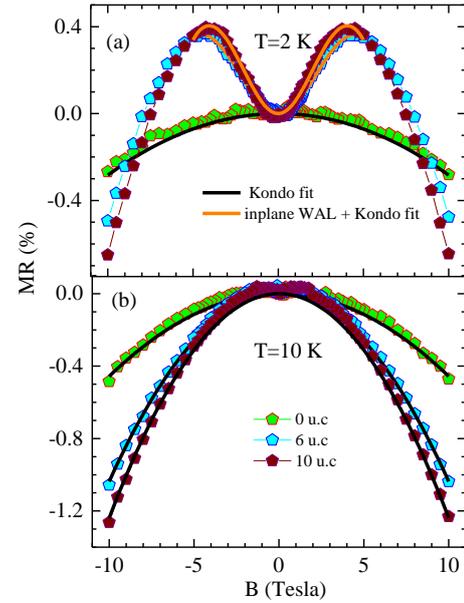

Figure 8. (a) $MR_\parallel$ of the δ=0, 6 and 10 u.c samples at 2 K.. All the samples show negative MR at higher fields which is attributed to Kondo effect. The δ=6 and 10 samples show positive MR at low fields followed by negative MR at higher fields. The positive MR is attributed to in-plane WAL effect. The black solid line represents Kondo fit (Eqn. (3)) to the δ=0 sample. The orange solid line represents in-plane WAL + Kondo fit (Eqn. (3 a) + (4)) to the δ=6 and 10 sample. The excellent fit indicates the presence of both WAL and Kondo effect in these samples at low fields. (b) $MR_\parallel$ of the δ=0, 6 and 10 u.c films at 10 K. The observed negative MR is due to Kondo effect. The black solid line represents Kondo fit using Eqn. (3).

Eqn. $3(a)$ is used to fit the experimental data in figure 8 (a),(b) (black solid line). The Kondo fit shows good agreement with our experimental data at 10 K and 2 K (δ=0 film), and also $R_K \propto \delta \propto 1/n_e$ at low temperatures, which conforms to the Kondo picture [61]. The fit parameters are shown in Table II. This result supports the presence of Kondo scattering at low temperatures. The origin of Kondo scattering in the LTO/STO (δ=0) can be attributed to the presence of localized polaronic unpaired electrons (S=1/2) that act as magnetic scattering centers to the delocalized electron gas at the interface [62]. However in doped samples (δ=6, 10), just the interaction between the delocalized and localized unpaired electrons is not enough to describe the Kondo scattering. If it were the case then $MR_∥$ should decrease with increase in thickness of the δ layer, as $n_e$ decreases. However, $MR_∥$ increases with δ layer thickness. Kondo behaviour has been reported in many Ce bulk systems [63-67]. Thus the enhancement in the Kondo effect is ascribed to the presence of $Ce^{3+}$ ions.

$MR_∥$ for δ=6 and 10 u.c samples at 2 K presents an interesting case, wherein the superposition of two scattering mechanism is manifested viz. Kondo and WAL effect. At lower fields, WAL effect dominates resulting in a positive slope in $MR_∥$, beyond 4 T, Kondo effect takes over which leads to a negative slope and eventually to negative $MR_∥$. Al'tshuler et al [68] were the first to study

Table II. Fitting parameters extracted from the Kondo fit to the in-plane MR data at 10K.

| Sample | $R_0(Ω/□)$ | $R_K(Ω/□)$ | $B_1$(T) |
|---|---|---|---|
| δ=0 u.c | 435 | 212 | 39.8 |
| δ=6 u.c | 515 | 1147 | 38.3 |
| δ=10 u.c | 557 | 1701 | 36.6 |

WAL effect in films with magnetic fields applied parallel to the film plane. The WAL contribution to $MR_∥$ is given by [68][69]:

$$\frac{\Delta R_□(B)}{[R_□(0)]^2} = \frac{e^2}{2\pi^2 \hbar} \ln(1 + \beta \frac{ed^2}{4\hbar B_\varphi} B^2) \quad (4)$$

Where, $d$ is the film thickness, $H_\varphi$ is the dephasing field and $\beta$ is a parameter that depends on the ratio of mean free path ($l_e$) and the film thickness. The ratio $l_e/d$ characterizes the parallel field transport into three regimes, viz. Altshuler-Aronov (AA) regime, where $l_e \ll d, \beta = 1/3$, Dugaev-Khmelnitskii (DK) regime, where $l_e \gg d, \beta = \frac{1}{16}\frac{d}{l_e}$ and Beenakker-van Houten (BvH) regime, where $l_e$ is comparable to d and $\beta < 1/3$. The mean free path is given by, $l_e = \hbar\sqrt{2\pi n_e}\mu/e$, $n_e$ and $\mu$ are the carrier density and mobility obtained from Hall measurements. For δ=6 and 10 u.c samples at 2 K, $l_e = 92$ Å and 75 Å which is comparable to the electron confinement in the STO side (12 nm) [7], thus $\beta$ lies in the BvH regime. In order to justify the contribution of both Kondo and WAL effect to $MR_∥$ for δ=6 and 10 u.c, we fit the experimental data by adding eqn. 3($a$) and (4). The excellent fit (solid orange line) in figure 8(a) supports our argument. The obtained fit parameters are shown in Table III.

Table III: Fitting parameters obtained from WAL+Kondo fit to the in-plane MR data at 2 K

| Sample | $R_0(Ω/□)$ | $R_K(Ω/□)$ | $B_1$(T) | $\beta$ |
|---|---|---|---|---|
| δ=0 u.c | 435 | 212 | 50.6 | - |
| δ=6 u.c | 515 | 1147 | 16.8 | 0.21 |
| δ=10 u.c | 557 | 1701 | 14.54 | 0.14 |

It was observed in our δ doped LTO/STO system that the SOI can be tuned by the δ layer thickness. SOI at the interface of STO based 2DEG can also be modulated by applying a back gate voltage ($V_g$). Positive $V_g$ tends to increase $n_e$ whereas a negative $V_g$ lowers $n_e$. Liang et al [56] reported a non-monotonic change in the SOI of LAO/STO and LVO/STO as a function of $V_g$; on increasing positive $V_g$, SOI first increases goes through a maximum and then decreases again. A large SOI has been predicted at the crossover region of $d_{xy} - d_{xz/yz}$ subbands (SOI is minimal at $d_{xy}$ and $d_{xz/yz}$ subbands), where orbital mixing induces large orbital angular momentum [70]. This explains the observations of Liang et al [56], where the increasing $n_e$ first populates the lower energy $d_{xy}$ subband followed by the higher $d_{xz/yz}$ subbands and in the process Fermi energy ($E_F$) crosses the $d_{xy} - d_{xz/yz}$ interface leading to enhanced SOI. Contrary to the experimental observations of Liang et al [56], we observe an increase in SOI with decreasing $n_e$ (increasing δ layer thickness). We attempt to explain our observations based on the screening effect of the interface potential. As the thickness of the δ layer increases, the (LaO)$^+$ planes which provide the confining potential for the charge carriers at the interface is further isolated from the STO surface. Furthermore, at lower temperatures the dielectric constant $\varepsilon$ of STO increases by two orders of magnitude [46]. These two effects strongly screen the confining potential at the interface, as a result of which the electron density spreads away from the interface, into the bulk of STO. As the confinement potential decreases the splitting of $d_{xy}$ and $d_{xz/yz}$ subbands also decrease [71]. At the interface, $d_{xy}$ subband has lower energy; however, away from the interface, in the bulk of STO the energies for the $d_{xz/yz}$ subbands are lower and hence are easily populated by the electrons [72][73]. Thus, the enhanced $\varepsilon$ of STO at low temperatures and increasing thickness of the δ layer facilitates the spread of electrons away from the interface, where it starts occupying $d_{xz/yz}$ subbands; this results in the crossover of $E_F$ from $d_{xy}$ to $d_{xz/yz}$ subbands leading to observed non-monotonic behaviour in SOI. Our arguments can be further supported if we look at the $k$-dependence of SOI of the $d_{xz/yz}$ subbands. $d_{xz/yz}$

subbands possesses only *k*-cubic dependence on SOI [74]. Furthermore, HLN theory accounts for only *k*-cubic SOI [56]. Hence the excellent fitting of WAL effect obtained using the HLN theory in our δ doped samples confirms the presence of only *k*-cubic SOI in our 2DEG system, which in turn implies the contribution of electrons in the $d_{xz/yz}$ subbands to the MR for the δ doped samples.

## 4. Conclusions

In summary, we have successfully deposited epitaxial thin films of CTO on $TiO_2$ terminated STO substrates. The CTO/STO films show an insulating behaviour which is attributed to the absence of 2DEG formation at the interface. XPS studies carried out on CTO/STO film show majority of the Ce and Ti to be in the 3+ valence state, although some amount of $Ti^{4+}$ and negligible amount of $Ce^{4+}$ are also present. The ratio of Ce : Ti (calculated from the Ce $3d_{5/2}$ and Ti $2p_{3/2}$ peak intensity) comes out to be 0.5, which indicates Ce vacancies ($V_{Ce}$) in the CTO lattice. We ascribe the presence of $Ti^{4+}$ ions in CTO to $V_{Ce}$, which incidentally results in the lack of 2DEG at the interface. $V_{Ce}$ also explains the decrease in the out-of plane lattice parameter of CTO obtained from XRD. M(T) measurements on CTO/STO film exhibits diamagnetic behaviour, which is different from the antiferromagnetic behaviour reported in the bulk CTO. We have also deposited δ unit cells of CTO at the LTO/STO interface. It has been observed that the electron density at the interface decreases with δ layer thickness. This provides strong evidence in favour of polar reconstruction being responsible for 2DEG formation at LTO/STO interface. At low temperatures, magneto transport measurement reveals the presence of Kondo scattering, EEI and SOI in the LTO/CTO/STO heterostructure. The strength of Kondo scattering increases with δ layer thickness. The increase in Kondo scattering is attributed to $Ce^{3+}$. SOI displays a non-monotonic behaviour with δ layer thickness, which is related to the screening of the electron confining potential at the interface. The decrease in confining potential decreases the splitting between the $d_{xy}$ and $d_{xz/yz}$ subbands and spreads 2DEG deeper into the STO bulk, where it starts to populate the Ti $d_{xz/yz}$ subbands. As $E_F$ crosses over to the $d_{xz/yz}$ subbands, a large SOI is observed at the crossover region of $d_{xy} - d_{xz/yz}$ where orbital mixing induces a large orbital momentum. Our results show that CTO δ layer can be used as an alternative to back gate voltage for SOI modulation in STO based 2DEG.

**Acknowledgements**

The authors would like to thank Prof. R. C. Budhani for the useful discussions. Financial assistance from IIT Kanpur is gratefully acknowledged.